# How much research output from India gets social media attention?

*Sumit Kumar Banshal, Vivek Kumar Singh\*, Pranab K. Muhuri and Philipp Mayr*

*Scholarly articles are now increasingly being mentioned and discussed in social media platforms, sometimes even as pre- or post-print version uploads. Measures of social media mentions and coverage are now emerging as an alternative indicator of impact of scholarly articles. This article aims to explore how much scholarly research output from India is covered in different social media platforms, and how similar or different it is from the world average. It also analyses the discipline-wise variations in coverage and altmetric attention for Indian research output, including a comparison with the world average. Results obtained show interesting patterns. Only 28.5% of the total research output from India is covered in social media platforms, which is about 18% less than the world average. ResearchGate and Mendeley are the most popular social media platforms in India for scholarly article coverage. In terms of discipline-wise variation, medical sciences and biological sciences have relatively higher coverage across different platforms compared to disciplines like information science and engineering.*

THE rapid growth of the Internet and social media has not only transformed businesses, organizations and society, but has also changed the entire process of scholarly information processing, including article storage, access and dissemination. Scholarly articles are now mentioned or shared on different social media platforms such as ResearchGate[1], Twitter[2], Facebook[3], Academia[4] and Mendeley[5]. Social media coverage and transactions regarding scholarly articles have become so popular that a new range of metrics has been developed, called altmetrics (for alternative metric), to measure and characterize social media coverage and transaction patterns[6,7]. Altmetrics is now an interesting area of study, where researchers analyse the social media coverage and consumption of scholarly articles, and sometimes even use them to predict the future citations of scholarly articles.

Several previous studies have explored different aspects of altmetrics, ranging from altmetrics-citation correlations to country-/region-specific studies. Some of these studies aimed to demonstrate if social media platforms can be used (or not) as a tool to attract more attention towards a published work[6,8–11]. Some other studies explored whether altmetrics could correlate with citations[12–14], with a few going to the extent to examine if it can complement citations or not[15]. There have also been studies to predict early citations from different social media platforms, such as Mendeley[16], ResearchGate and Google Scholar[17], altmetric.com[18], CiteULike bookmarks[19], etc.

In addition to studies that focus on interconnections between altmetrics and traditional scientometrics, several studies focused on country-/region-specific altmetric analysis of scholarly articles, though they are limited in number. A study focused on evaluating research work from Taiwan using 18 different online media-based indicators[20]. Cho[21] performed a study on articles from South Korea and found that Mendeley had more correlation with traditional impact than other social media platforms. Holmberg and Woo[22] studied the online platform visibility for scientific journals from South Korea. Bangani[23] studied institutional repositories of South Africa, and assessed the impact of theses and dissertations using altmetrics. Some other studies mapped Chinese scholarly performance using different online platforms such as Twitter[24], and Twitter and Mendeley[25]. Wang et al.[26] pointed out that less visibility of Chinese papers in social media platforms is due to less accessibility of public platforms in China. Teixeira de Silva[27] discussed the limitations of research policies in China about social media indices. Some studies also did cross-country or regional comparisons on alternative metrics. A study compared higher education institutes from the US and Europe on their ResearchGate visibility and ResearchGate (RG) scores[28].

Sumit Kumar Banshal and Pranab K. Muhuri are in the Department of Computer Science, South Asian University, New Delhi 110 021, India; Vivek Kumar Singh is in the Department of Computer Science, Banaras Hindu University, Varanasi 221 005, India; Philipp Mayr is in the GESIS Leibniz Institute for Social Sciences, Cologne, Germany.
\*For correspondence. (e-mail: vivek@bhu.ac.in)

There are, however, no previous studies using altmetrics for research articles from India, except one[29], wherein the authors have analysed the ResearchGate platform for coverage and disciplinary variations of research articles from India. However, they only worked with ResearchGate data and did not compare the social media coverage values of Indian research output with the world average. The present study fills this gap and also answers several other important research questions. A detailed and systematic analysis of altmetric attention of scholarly articles from India in several popular social media platforms like Twitter, Facebook, Mendeley, News, etc. is performed. The study also measures coverage levels and compares them with the world average, and identifies disciplinary variations in coverage of scholarly articles from India in different social media platforms. More precisely, the study aims to answer the following questions:

RQ1: How much scholarly research output from India is covered, mentioned and discussed in social media platforms?
RQ2: Is the social media coverage of research output from India is at par or below/above the world average?
RQ3: How are different disciplines distributed in research output from India and the world, as indexed in the Web of Science (WoS) and also as found in altmetric.com?
RQ4: Which social media platforms are more popularly used by Indian authors/researchers?
RQ5: Are there any discipline-wise variations in social media coverage and consumption patterns of scholarly articles from India, and how similar or dissimilar are they with respect to the worldwide pattern?

**Data and methodology**

Data for the study was obtained from two sources: WoS[30] and altmetric.com[31]. First, data from WoS were downloaded for research publications originating from India (i.e. those having at least one author affiliated to an Indian institution) for the year 2016. A total of 91,106 publication records were found for India, out of which 88,259 records were unique. Among these 88,259 records, 76,621 were found to have digital object identifier (DOI). WoS data were collected during 5–10 May 2017. For each downloaded record, standard dataset fields like title, authors, publication type, publication source, citations, references, etc. were obtained. In order to compare the altmetric coverage of India with the world average, data for the whole world for the corresponding year were also downloaded. A total of 2,528,868 publication records were found for the whole world, out of which 1,460,124 were found to have DOI.

Secondly, altmetric data were obtained for publication records from WoS through a DOI look-up in the altmetric.com website. The altmetric data downloaded were updated till 22 May 2018. Out of 76,621 records found in WoS for India having DOI, a total of 21,644 (approx. 28.5%) were found covered in altmetric.com. Similarly, for the world, out of 1,460,124 records from WoS with DOI, a total of 681,274 (approx. 47%) were found to be covered in altmetric.com. The altmetric.com website is a data repository which collates attentions and mentions about scholarly articles from a wide range of online networks and media. It provides 18 different types of online data, including from different networks like Twitter, Facebook, Weibo, Google Plus, LinkedIn; online news, blog-sites; news and information aggregators like Reddit, Pinterest; academic networks like Mendeley, F1000 and online encyclopaedia like Wikipedia. It also generates an aggregated score for each scholarly article by combining data from different platforms.

Some data from the ResearchGate platform are also shown for comparison with results of the present study. The data for research papers from India indexed in WoS were obtained using a web crawler, originally written for an earlier work[29]. For each record in WoS, the ResearchGate platform was searched by the crawler to extract relevant data. The extracted data were then analysed by a computational procedure that comprised of several codes written in R programming language. The results obtained are shown in tabular form for better understanding.

The data downloaded, as above, were analysed using computational data analysis. While coverage levels could be computed easily, the disciplinary variation result computation required tagging each publication record into specific discipline(s). For this, each publication record in the dataset was tagged into one of the 14 broad research disciplines, as proposed in an earlier work[32]. This tagging was done using Web of Science Category (WC) field information in publication records. One record can be tagged with multiple disciplines of research based on its WC entries. The 14 broad disciplines in which the publication records were tagged are as follows: agriculture (AGR), arts and humanities (AH), biology (BIO), chemistry (CHEM), engineering (ENG), environment science (ENV), geology (GEO), information sciences (INF), material science (MAR), mathematics (MAT), medical science (MED), multidisciplinary (MUL), physics (PHY) and social science (SS). Thus the 255-category division of articles in WoS was reduced to these 14 broader disciplines and each publication record was tagged with one (or in some cases more) broad discipline. All further analysis on disciplinary variations in altmetric coverage was done across these 14 broad disciplines. A set of computational processes was written in R programming language to process the data and obtain analytical results.

## Results

The computational analysis of data produced analytical results about altmetric coverage level of research output from India, its comparison with the world average, disciplinary distribution of data and disciplinary variations in altmetric coverage.

*Coverage*

The first analytical result obtained shows overall altmetric coverage of research output from India. Out of total 76,621 research papers published from India as indexed in WoS with DOI, only 21,644 are found to be included in altmetric.com, i.e. 28.5% of the research output from India is covered in social media platforms. Coverage in altmetric.com indicates coverage by social media platforms like Mendeley, Twitter, Facebook, etc. Table 1 shows data for coverage of research articles from India in different social media platforms. It can be observed that Mendeley has the highest coverage of 27.2%, followed by Attention score with 24.1%, Twitter with 21.6% and Facebook with 4.7%.

It would be interesting to compare altmetric coverage of research output from India with the world average. For the whole world, out of total 1,460,124 research papers found indexed in WoS with DOI, 681,274 are found to be covered by altmetric.com, i.e. approximately 47% of research output from the world is covered in some social media platform. Table 2 shows the platform-wise coverage data for the world research output. It can be observed that Mendeley has highest coverage of 43.5%, followed by Attention score with 37.1%, Twitter with 34.4% and Facebook with 9.1%. Thus, it is evident that research output from India, in general, is getting lesser attention in social media platforms compared to the world average. Figure 1 shows the ratio of data for different platforms for India and the world. It can be seen that India's share of altmetric coverage to the world lies between 2% and 5% for different platforms. This may be related to the country's contribution to annual publication data indexed in WoS, which is approx. 5%.

**Table 1.** Coverage of articles from India in different social media platforms as captured by altmetric.com

| Mention type | TP | Percentage (%)* | Total mention | Mention/paper |
|---|---|---|---|---|
| Mendeley | 20,815 | 27.2 | 353,817 | 16.998 |
| Attention score | 18,449 | 24.1 | 136,222 | 7.384 |
| Twitter | 16,569 | 21.6 | 102,176 | 6.167 |
| Facebook | 3,594 | 4.7 | 6,960 | 1.937 |
| News Mentions | 1,455 | 1.9 | 9,528 | 6.548 |
| Blog | 949 | 1.2 | 1,892 | 1.994 |
| Google | 517 | 0.7 | 1,695 | 3.279 |
| Wiki | 496 | 0.6 | 760 | 1.532 |
| Reddit | 229 | 0.3 | 270 | 1.179 |
| Policy | 157 | 0.2 | 229 | 1.459 |
| Peer review | 149 | 0.2 | 315 | 2.114 |
| F1000 | 137 | 0.2 | 151 | 1.102 |
| Patent | 68 | 0.1 | 80 | 1.176 |

*With respect to total papers for 2016 for India indexed in Web of Science (WoS) with DOI = 76,621.

**Table 2.** Coverage of articles from the World in different social media platforms as captured by altmetric.com

| Mention type | Total papers | Percentage (%)* | Total mentions | Mention/paper |
|---|---|---|---|---|
| Mendeley | 634,825 | 43.5 | 17,743,006 | 27.949 |
| Attention score | 542,363 | 37.1 | 6,813,120 | 10.001 |
| Twitter | 501,833 | 34.4 | 4,441,526 | 8.851 |
| Facebook | 133,439 | 9.1 | 308,801 | 2.314 |
| News Mentions | 69,261 | 4.7 | 533,952 | 7.709 |
| Blog | 46,802 | 3.2 | 91,557 | 1.956 |
| Google | 21,108 | 1.4 | 48,986 | 2.321 |
| Wiki | 13,674 | 0.9 | 20,521 | 1.501 |
| Reddit | 9,273 | 0.6 | 12,462 | 1.344 |
| Policy | 9,244 | 0.6 | 12,497 | 1.352 |
| Peer review | 3,027 | 0.2 | 5457 | 1.803 |
| F1000 | 7,025 | 0.5 | 8148 | 1.16 |
| Patent | 6,309 | 0.4 | 9515 | 1.508 |

*With respect to total papers for 2016 for world in WoS = 1,460,124.

*Discipline-wise distribution of data*

Before we examine the discipline-wise variations in coverage of articles in altmetrics, it would be interesting to see disciplinary distribution of the whole data downloaded from WoS as well as those found in altmetric.com for both India and the world. Figure 2 shows the distribution of papers from different disciplines in total research output from India as indexed in WoS (left) and as found in altmetric.com (right). There are interesting patterns in the figure. MED accounts for 29.6% papers in WoS, whereas in altmetric.com it accounts for 39.7% of total papers. Similarly, BIO shows 11.3% contribution to research output indexed in WoS, but in altmetric.com it shows 18%. Thus, there are disciplines which are proportionately covered more in altmetric.com than in WoS. These are MED, BIO, ENV, SS, etc. Some disciplines have less proportionate contribution in altmetric.com than WoS, i.e. PHY, ENG and INF. PHY has 18.6% papers in WOS, and 14.1% in altmetric.com. ENG has 12.7% papers in WoS, but only 5.4% in altmetric.com. INF has 5.1% papers in WOS, and 2.3% in altmetric.com. Therefore, it is clearly seen that some disciplines (such as MED, BIO) attract more social media coverage than their publication volume compared to other disciplines (such as ENG, INF).

In order to compare these trends with the world average, WoS and altmetric discipline-wise distribution for the world data was also obtained. Figure 3 shows the proportionate contribution of different disciplines in total research output from the world, as indexed and distributed in WoS (left) and as found in altmetric.com (right). For the world data, similar differences in distributions are found. However, in this case some disciplines, which are proportionately more covered in Indian data show less contribution. ENV is the most noticeable, having 4.8% papers in WoS, but lesser contribution (4.5%) in altmetric.com. Patterns for most other disciplines are, however, similar (such as MED with 30.2% and 41.4%, BIO 8.4% and 11.9%, ENG 5.2% and 2.3% in WOS and altmetric.com respectively). Thus, more altmetric attention to research output from some disciplines is a common pattern in India and the world, with a few exceptions.

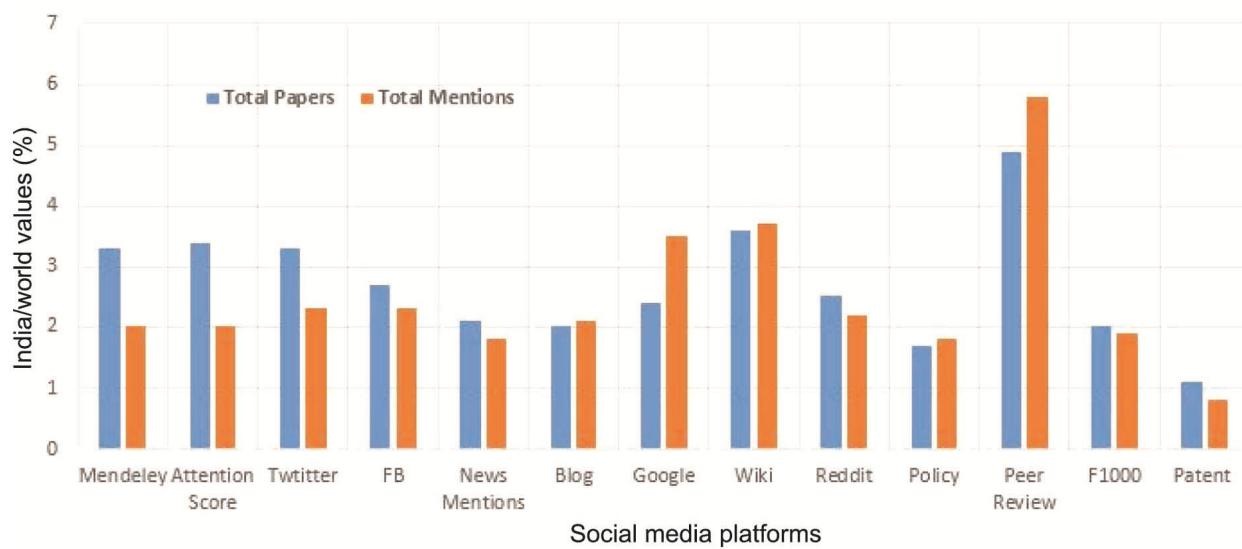

**Figure 1.** Altmetric coverage levels of India divided by the world average.

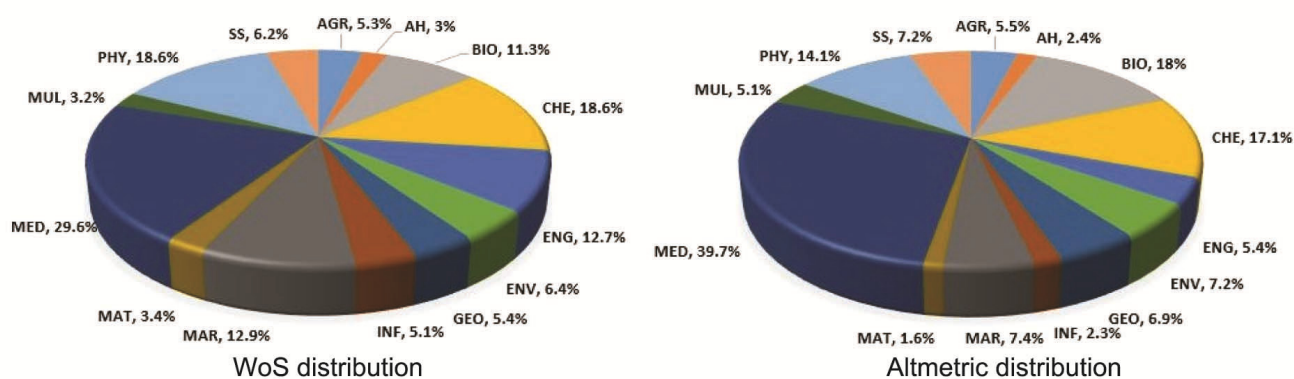

**Figure 2.** Discipline-wise article distribution in the Web of Science (WoS) and altmetric.com for publications from India.

*Discipline-wise variations in coverage*

It would be relevant to also find discipline-wise differences in overall altmetric coverage of articles from India and compare them with the world. Table 3 shows data for altmetric coverage of research output from India in four different social media platforms, viz. Twitter, Facebook, News Mention and Mendeley. It can be observed that in Twitter, BIO, MUL and MED have the highest coverage, with values of 38.3%, 37.3% and 30.7% respectively. ENG and INF have least coverage of 5.9% and 7.1% respectively. In Facebook, MUL has the highest coverage of 11% followed by SS with 8.4%, MED with 7.9% and AGR with 7%. ENG and INF again have the lowest coverage of 1% and 1.1% respectively. In News Mention, MUL have highest coverage of 6.6%, followed by BIO with 4.3%. MAT has the least coverage of 0.3% followed by ENG with 0.4%. In Mendeley, BIO has the highest coverage of 44.4%, followed by MUL with 44.3% and MED with 36.4%, ENG has the least coverage of 11.6%, followed by MAT with 12.1%. Thus, it is clearly seen that disciplines like MUL, BIO and MED have higher altmetric coverage percentage across platforms, and disciplines like ENG, INF and MAT have lesser altmetric coverage. In terms of platforms, Mendeley and Twitter have in general higher coverage than Facebook and News Mention.

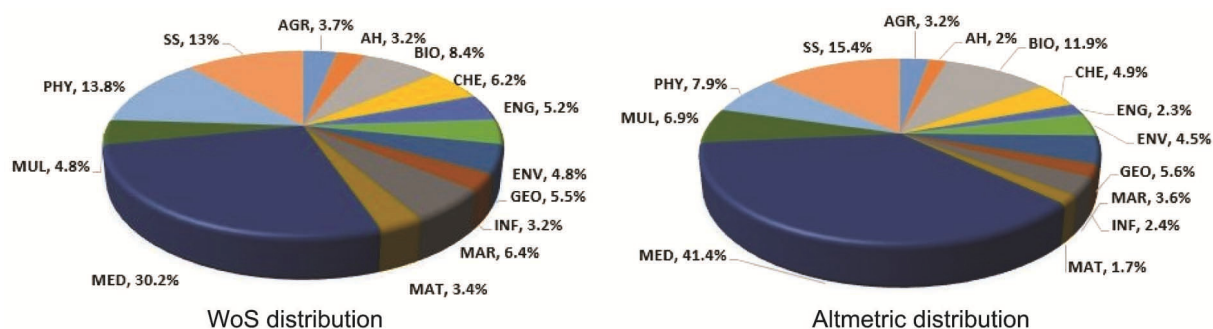

**Figure 3.** Discipline-wise article distribution in WoS and altmetric.com for publications from the World.

**Table 3.** Discipline-wise variations in coverage in different social media platforms for publications from India

| Discipline | Articles in WoS | Twitter | | Facebook | | News Mentions | | Mendeley | |
|---|---|---|---|---|---|---|---|---|---|
| | | No. of articles | Coverage percentage | No. of articles | Coverage percentage | No. of articles | Coverage percentage | No. of articles | Coverage percentage |
| AGR | 4,099 | 840 | 20.5 | 286 | 7 | 79 | 1.9 | 1,159 | 28.3 |
| AH | 2,318 | 356 | 15.4 | 79 | 3.4 | 15 | 0.6 | 463 | 20 |
| BIO | 8,626 | 3,307 | 38.3 | 569 | 6.6 | 374 | 4.3 | 3,832 | 44.4 |
| CHE | 14,270 | 2,937 | 20.6 | 392 | 2.7 | 181 | 1.3 | 3,571 | 25 |
| ENG | 9,694 | 569 | 5.9 | 98 | 1 | 35 | 0.4 | 1,129 | 11.6 |
| ENV | 4,930 | 1,151 | 23.3 | 254 | 5.2 | 93 | 1.9 | 1,533 | 31.1 |
| GEO | 4,105 | 1,053 | 25.7 | 149 | 3.6 | 73 | 1.8 | 1,465 | 35.7 |
| INF | 3,890 | 277 | 7.1 | 44 | 1.1 | 20 | 0.5 | 492 | 12.6 |
| MAR | 9,856 | 1,110 | 11.3 | 152 | 1.5 | 49 | 0.5 | 1,536 | 15.6 |
| MAT | 2,641 | 224 | 8.5 | 31 | 1.2 | 9 | 0.3 | 319 | 12.1 |
| MED | 22,676 | 6,955 | 30.7 | 1,785 | 7.9 | 690 | 3 | 8,258 | 36.4 |
| MUL | 2,472 | 922 | 37.3 | 271 | 11 | 164 | 6.6 | 1,096 | 44.3 |
| PHY | 14,255 | 2,250 | 15.8 | 293 | 2.1 | 120 | 0.8 | 2,857 | 20 |
| SS | 4,729 | 1,256 | 26.6 | 395 | 8.4 | 97 | 2.1 | 1,515 | 32 |

Table 4 shows the equivalent results for the world data. In Twitter, MUL has the highest coverage of 55.7%, followed by BIO with 54.6% and MED with 50.8%. ENG and INF have the least coverage of 7.5% and 9.3% respectively. In Facebook, MUL, MED and BIO have the highest coverage of 17.8%, 16% and 13.7% respectively. INF and ENG have the least coverage of 1.3% and 1.4% respectively. In News Mention, MUL, MED and BIO have the highest coverage of 13%, 7.5% and 7.3% respectively. ENG and INF have the least coverage of 0.5% and 0.8% respectively. In Mendeley, MUL, BIO and MED have the highest coverage of 63.6%, 62.1% and 60% respectively. ENG and MAT have the least coverage of 18.6% and 19.8% respectively. Here, INF has relatively better coverage of 30.5%, compared to the Indian pattern. In terms of platforms, Mendeley and Twitter have overall higher coverage than Facebook and News Mention.

Thus, it is interesting to observe that there are similar discipline-wise variations in the coverage of different disciplines in research output from India and the world. MUL, BIO and MED, in general, have more social media visibility, while ENG, MAT and INF have relatively less social media visibility. Mendeley coverage level of INF for the world data is an exception recorded. Similarly, in both cases, Mendeley and Twitter have higher coverage percentage while Facebook and News Mention have lower coverage percentage. However, coverage percentage for research output from India is significantly lesser than that for the world data, across disciplines. Figure 4 shows the ratio of India's altmetric coverage vis-à-vis the World data for different disciplines for the four social platforms. The coverage ratio varies from approximately 2% to 12% for different disciplines. Thus, we can conclude that not all disciplines attract similar amount of social media attention.

### Coverage and discipline-wise variations in ResearchGate

This study also presents some analytical results from the ResearchGate platform, which are not covered in altmetric.com data. Table 5 shows some important results about coverage for different disciplines in the ResearchGate platform. It can be observed that coverage percentage for almost all disciplines is higher compared to other social media platforms seen earlier. Unlike, other platforms, here GEO has highest coverage of 94.9% followed by MUL with 80.2%, AGR with 75.2% and SS with 74.3%. Thus, the patterns of coverage of different disciplines in ResearchGate are different from other platforms. The table also shows data about reads and citations in ResearchGate for different disciplines. It may, however, be noted that comparison with corresponding data from the world could not be presented due to huge volume of such data (and hence longer-time requirement to crawl the Web for the same).

Table 4. Discipline-wise variations in coverage in different social media platforms for publications from around the world

| Discipline | Articles in WoS | Twitter | | Facebook | | News Mentions | | Mendeley | |
|---|---|---|---|---|---|---|---|---|---|
| | | No. of articles | Coverage percentage | No. of articles | Coverage percentage | No. of articles | Coverage percentage | No. of articles | Coverage percentage |
| AGR | 53,749 | 16,132 | 30 | 4,406 | 8.2 | 1,468 | 2.7 | 20,784 | 38.7 |
| AH | 47,186 | 8,690 | 18.4 | 2,025 | 4.3 | 350 | 0.7 | 10,763 | 22.8 |
| BIO | 123,180 | 67,281 | 54.6 | 16,850 | 13.7 | 9,006 | 7.3 | 76,480 | 62.1 |
| CHE | 90,959 | 24,733 | 27.2 | 4,673 | 5.1 | 2,332 | 2.6 | 31,331 | 34.4 |
| ENG | 75,834 | 5,663 | 7.5 | 1,067 | 1.4 | 355 | 0.5 | 14,128 | 18.6 |
| ENV | 69,709 | 22,196 | 31.8 | 4,722 | 6.8 | 2,219 | 3.2 | 28,961 | 41.5 |
| GEO | 80,477 | 26,873 | 33.4 | 5,445 | 6.8 | 3,599 | 4.5 | 35,902 | 44.6 |
| INF | 46,438 | 4,330 | 9.3 | 583 | 1.3 | 373 | 0.8 | 14,151 | 30.5 |
| MAR | 94,117 | 15,096 | 16 | 2,508 | 2.7 | 1,674 | 1.8 | 23,280 | 24.7 |
| MAT | 49,385 | 5,773 | 11.7 | 792 | 1.6 | 618 | 1.3 | 9,777 | 19.8 |
| MED | 441,032 | 224,132 | 50.8 | 70,401 | 16 | 33,021 | 7.5 | 264,405 | 60 |
| MUL | 69,445 | 38,675 | 55.7 | 12,371 | 17.8 | 9,021 | 13 | 44,194 | 63.6 |
| PHY | 201,373 | 33,571 | 16.7 | 5,973 | 3 | 3,908 | 1.9 | 50,031 | 24.8 |
| SS | 189,835 | 78,799 | 41.5 | 24,557 | 12.9 | 9,258 | 4.9 | 96,180 | 50.7 |

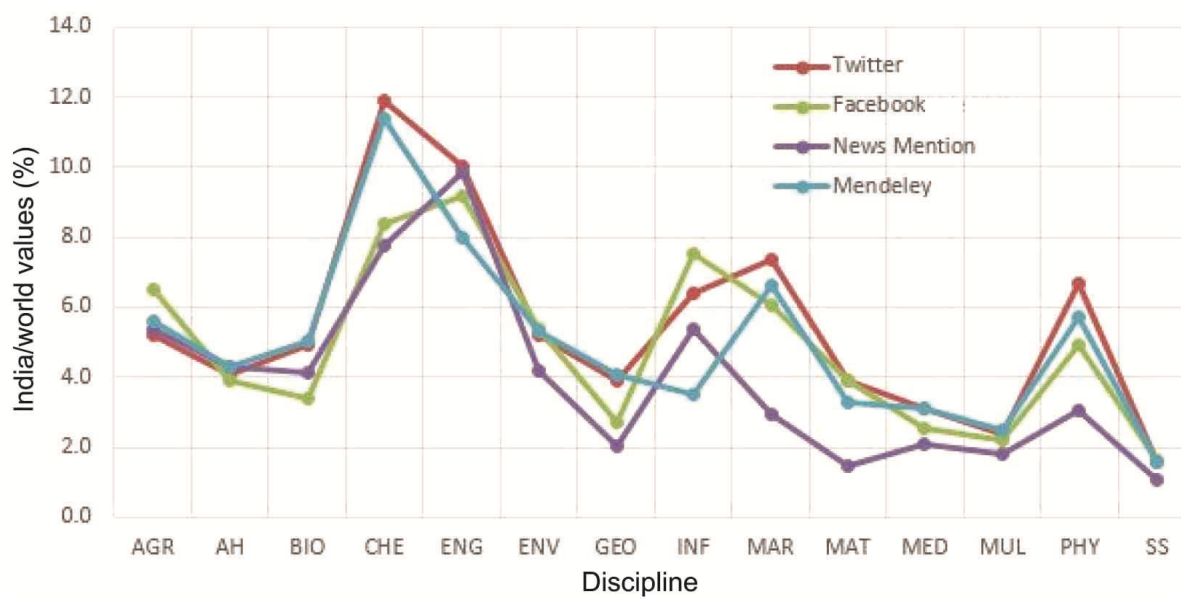

Figure 4. Altmetric coverage of India divided by world average for different disciplines.

## Conclusion

This study presents interesting results about social media coverage of research output from India and how it compares to the world average. Discipline-wise differences in data distribution and coverage are also obtained and analysed. The results obtained successfully answer the research questions proposed here. Analytical results show the following five important outcomes: First, the overall social media coverage of research output from India is 28.5%, which is significantly lower than the world average of 46.7%. Second, some disciplines like MED, BIO and MUL get more social media coverage compared to their publication volume, whereas some other disciplines like ENG, INF and MAT get less social media coverage in comparison to their publication volume. Third, there exist, discipline-wise variations in social media coverage of research output from India, which are in general similar to the worldwide pattern, with a few exceptions. Fourth, Mendeley and Twitter platforms have in general higher coverage of research output from India as well as the world compared to Facebook and News Mention. Interestingly, Facebook has the largest number of users from India. Further, ResearchGate platform has significantly higher coverage for Indian data compared to the other platforms. Fifth, overall coverage and pattern of discipline-wise variations in coverage in ResearchGate platform for research output from India are quite different from the other platforms. Here, GEO, MUL and AGR are the most covered, while CHEM and PHY are the least covered disciplines. It appears that research output from India in general is either less connected directly to societal concerns and hence less covered in social media platforms, or that lower coverage may be because of lesser penetration/academic usage of social media platforms in the country by researchers/readers in these areas.

Table 5. Discipline-wise reading-citation patterns in ResearchGate

| Discipline | Articles in WOS | Articles in RG | Coverage percentage | Reads/paper | Papers with at least one citation | Percentage of cited papers |
|---|---|---|---|---|---|---|
| AGR | 4,099 | 3,081 | 75.2 | 61.59 | 1,646 | 53.4 |
| AH | 2,318 | 1,580 | 68.2 | 61.54 | 924 | 58.5 |
| BIO | 8,626 | 6,238 | 72.3 | 61.22 | 3,647 | 58.5 |
| CHE | 14,270 | 8,960 | 62.8 | 59.54 | 5,558 | 62 |
| ENG | 9,694 | 6,587 | 67.9 | 60.79 | 3,955 | 60 |
| ENV | 4,930 | 3,393 | 68.8 | 71.09 | 2,046 | 60.3 |
| GEO | 4,105 | 3,895 | 94.9 | 66.37 | 2,292 | 58.8 |
| INF | 3,890 | 2,706 | 69.6 | 61.06 | 1,558 | 57.6 |
| MAR | 9,856 | 6,687 | 67.8 | 44.19 | 4,134 | 61.8 |
| MAT | 2,641 | 1,853 | 70.2 | 60.49 | 1,021 | 55.1 |
| MED | 22,676 | 16,167 | 71.3 | 54.45 | 7,734 | 47.8 |
| MUL | 2,472 | 1,982 | 80.2 | 65.26 | 1,113 | 56.2 |
| PHY | 14,255 | 9,377 | 65.8 | 54.54 | 5,660 | 60.4 |
| SS | 4,729 | 3,512 | 74.3 | 52.4 | 1,773 | 50.5 |

This study does not explore the causal/driving factors behind higher social media coverage of research output from some disciplines and relatively lower coverage for a few other disciplines, but it would definitely be an interesting exercise that will need further analysis. One possible factor may be that research output from MED, BIO and MUL gets higher social media coverage because these disciplines are more connected with the daily lives of common people compared to others like ENG, INF and MAT, which are more specialized and technical in nature, and hence may not be easily understood by common people. Further, much higher publication volume of MED, BIO and MUL and the large number of researchers producing these outputs could be another factor, as it results into higher chances of being covered in social media platforms. The publication sources for MED, BIO and MUL somehow have a relatively well-developed system of social media connections compared to journals in many other disciplines. However, a proper understanding of the factors needs further exploration and a more focused analysis of the worldwide data. This study also does not analyse any gender-related differences in social media coverage of scholarly articles from India, which can be taken up as an extension of the present work.

ACKNOWLEDGEMENTS. We acknowledge the access provided to data of altmetric.com by Stacy Konkiel, Director of Research and Education at altmetric.com. We also acknowledge the enabling support provided by the Indo-German Joint Research Project titled 'Design of a sciento-text computational framework for retrieval and contextual recommendations of high-quality scholarly articles' (Grant No. DST/INT/FRG/DAAD/P-28/2017) for this work.

Received 14 May 2019; revised accepted 6 June 2019

doi: 10.18520/cs/v117/i5/753-760